\documentclass[aps,prb,preprint,groupedaddress]{revtex4}
\usepackage{times,xspace}
\usepackage{amsfonts}
\usepackage{amsmath}
\usepackage{amssymb}
\usepackage[final]{graphicx}
\begin{document}
\title{Shunt-capacitor-assisted synchronization of oscillations in intrinsic Josephson junctions stack}
\author{I. Martin, G\'{a}bor B. Hal\'{a}sz, L. N. Bulaevskii}
\affiliation{Los Alamos National Laboratory, Los Alamos, New Mexico 87545}
\author{and A. E. Koshelev}
\affiliation{Materials Science Division, Argonne National Laboratory, Argonne,
Illinois 60439}

\date{\today}

\begin{abstract}
We show that shunt capacitor stabilizes synchronized oscillations in intrinsic Josephson junction stacks biased by DC current.
This synchronization mechanism has an effect similar to the previously discussed radiative coupling between junctions, however, it is not defined by the geometry of the stack.
It is particularly important in crystals with smaller number of
junctions, where radiation coupling is week, and is comparable with the effect of strong super-radiation in crystal with many junctions. The shunt also helps to enter the phase-locked
regime in the beginning of oscillations, after switching on the bias current.
 Shunt may be used to tune radiation power, which drops as shunt capacitance
increases.
\end{abstract}
\maketitle

Recently THz radiation was obtained from mesa-type layered crystals with
intrinsic Josephson junctions (IJJ) \cite{Oz}. The number of junctions in that
case was not very large, about 600, and power of radiation was enhanced by
exciting resonance modes inherent to the crystal, which acts as a cavity. Part of
energy stored in an excited mode leaked outside the crystal as radiation. The limitation of
such a design is that the radiation frequency is fixed by the crystal resonances and
cannot be continuously tuned. A general design for high power tunable source of THz radiation
based on IJJ in layered superconductors was discussed in Ref.\onlinecite{BK},
see Fig.~1. The main idea behind this design is to get radiation from crystal
boundaries from many synchronized IJJ (up to $N=10000$) biased with the DC
current. The current induces DC voltage $V$ between neighboring layers and
thus produces Josephson oscillations with the frequency $\omega_J=2eV/\hbar$
tunable by the DC current.
It was proposed in Ref.~\onlinecite{BK} to use a crystal with dimensions
$L_y\gtrsim c/\omega_J\gg L_x$ to eliminate effect of resonance modes in the
$x$-direction and use metallic screens at $|z|>L_z/2$ to eliminate destructive
interference of electromagnetic waves coming from the surfaces $x=\pm L_x/2$.
When all junctions are synchronized, the radiation
power emitted from the crystal edge is proportional to $N^2$ and may reach
1 mW from crystal with $L_y=300$ $\mu$m, $L_x=4$ $\mu$m and $L_z=40$ $\mu$m.

\begin{figure}[ptb]
\begin{center}
\includegraphics[width=3.4in]{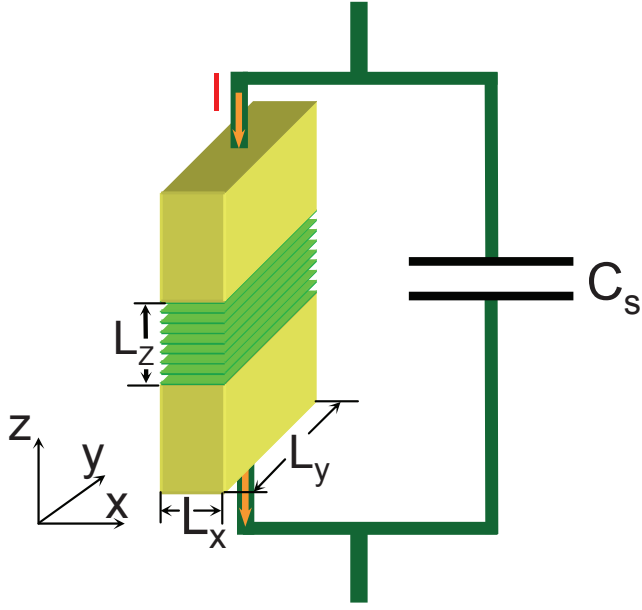}
\end{center}
\caption{Stack of intrinsic Josephson junctions shunted by external
capacitance. Light green plates are metallic screens, superconducting layers
are shown by dark green.}
\label{Fig-StackCapShunt}%
\end{figure}

It was shown in Ref.\ \onlinecite{BK} that radiation by the junctions itself can
synchronize oscillations.
Without radiation, in-phase oscillations in different
junctions may be unstable due to excitation of the Fiske modes in the layered crystal.
Here we propose additional mechanism of synchronization of Josephson oscillations
by means of external shunt capacitor, see Fig. \ref{Fig-StackCapShunt}. It works
in a way similar to radiation from the crystal,  but now all junctions
contribute to the electric field inside the shunt capacitor. The
effect of shunt stabilization of synchronized oscillations in an array of
point-like Josephson junctions was discussed previously by Chernikov and
Schmidt  \cite{CS}. Here we generalize their results for extended IJJ and find
stabilization condition for such systems. We calculate stabilization effects of
both, radiation and shunt, and compare them quantitatively. We show that
stabilization effects of shunt with moderate capacitance and of radiation in the
super-radiation regime (large number of junctions, of order $10^4-10^5$) are comparable, while
shunt capacitor is much more effective in keeping oscillations in different
junctions synchronized at smaller $N$. We demonstrate also that increase of shunt capacitance
results in suppression of radiation and thus radiation power may be tuned by shunt.

To account for the effect of external shunt on oscillations in the IJJ, we use the
Lagrangian approach formulated in Ref.~\onlinecite{BD}. The Lagrangian for the system with shunt
shown in Fig.~\ref{Fig-StackCapShunt} is
\begin{eqnarray}
&&{\cal L}\{\varphi_n\}=\frac{\Phi_0^2s}{16\pi^3\lambda_{ab}^2}\sum_n\int d{\bf
r}\left[\frac{1}{2c_0^2}(1-\alpha\nabla_n^2)^{-1}\dot{\varphi}_n^2-
\frac{1}{\lambda_J^2}(1-\cos\varphi_n)-\frac{1}{2}{\bf Q}_n^2
\right]- \nonumber\\
&&\int d{\bf r}dz\frac{({\rm curl} {\bf A})^2}{8\pi}
+\frac{\hbar^2}{8e^2}C_sN^2\dot{\varphi}^2.
\end{eqnarray}
Here $\varphi_n({\bf r},t)$ is the gauge-invariant phase difference between the
layers $n$ and $n+1$, the coordinates inside layer are ${\bf r}=x,y$, phase
difference gradients are
$\nabla\varphi_n=(\nabla_x\varphi_n,\nabla_y\varphi_n)$, the London penetration
lengths are $\lambda_c$ and $\lambda_{ab}$ for currents between layers and
inside layers, respectively, $\epsilon_c$ and $\epsilon_{ab}$ are
high-frequency dielectric constant for electric fields perpendicular to layers
(along the $z$-axis) and along layers, $\ell =\lambda_{ab}/s$, where $s$ is the
interlayer distance, $c_0=c/(\sqrt{\epsilon_c}\ell)$ and $\lambda_J=\gamma s$,
where $\gamma=\lambda_c/\lambda_{ab}$ is the anisotropy ratio, and
$\Phi_0=\pi\hbar c/e$.

The first two terms account for the electro-chemical and the Josephson energies of the IJJ. The factor $(1-\alpha\nabla_n^2)^{-1}$, with second
discrete derivative $\nabla_n^2A_n=A_{n+1}+A_{n-1}-2A_n$ and
$\alpha=e^{-1}s^{-2}(4\pi)^{-1}\partial\mu/\partial \rho$, originates from the relation between gauge invariant time derivative of the phase difference and difference in the
chemical potentials, see Ref.~\onlinecite{KT}:
\begin{equation}
\hbar\frac{\partial\varphi_n}{\partial
t}=e(V_n-V_{n+1})+\frac{\partial\mu}{\partial\rho}(\rho_n-\rho_{n+1}),
\label{charge}
\end{equation}
where $V_n$, $\rho_n=(E_{zn}-E_{z,n-1})/(4\pi s)$, and $\mu_n$ are the
potential, the charge density, and the chemical potential in the layer $n$, while ${\bf E}_n$
is the electric field in the junction $n$. Eq.~(\ref{charge}) results
in the relation
\begin{equation}
E_{zn}({\bf
r},t)=(1-\alpha\nabla_n^2)^{-1}(B_c\ell\lambda_c/c)\dot{\varphi}_n({\bf
r},t), \ \ \ B_c=\Phi_0/(2\pi\lambda_{ab}\lambda_c). \label{elec}
\end{equation}
The next terms in the square brackets  account for the kinetic energy of the intralayer currents. The intralayer current is ${\bf
j}_n=(c\Phi_0/8\pi^2\lambda_{ab}^2){\bf Q}_n$, where ${\bf
Q}_n=-\nabla\phi_n-(2\pi/\Phi_0){\bf A}_n$, where we introduced the phase $\phi_n$
of the superconducting order parameter and the vector potential ${\bf A}_n({\bf
r})$ in the layer $n$.
The fourth term in the Lagrangian is the energy of magnetic field inside
the crystal. The term with $\dot{Q}_n$ is omitted because its contribution is negligible at the low frequencies discussed here.

The electric energy inside the shunt capacitor $C_s$ is
accounted for by the last term in Lagrangian, written  in terms of the average phase
difference
\begin{equation}
\varphi(t)=\frac{1}{N}\sum_n\int\frac{d{\bf r}}{L_xL_y}\varphi_n({\bf r},t).
\end{equation}
The dissipative function is ${\cal R}\{\varphi_n\}={\cal
R}_c\{\varphi_n\}+{\cal R}_{ab}\{\varphi_n\}$, where
\begin{eqnarray}
&&{\cal R}_c\{\varphi_n\}=\frac{\Phi_0^2s}{32\pi^3\lambda_{ab}^2}\sum_n\int d{\bf r}\frac{4\pi\sigma_c}{c_0^2\epsilon_c}\dot{\varphi}_n^2, \\
&&{\cal R}_{ab}\{\varphi_n\}=\frac{\Phi_0^2s^3}{32\pi^3\lambda_{ab}^2}\sum_n\int
d{\bf r}\frac{4\pi\sigma_{ab}}{c_0^2\epsilon_{ab}}\dot{{\bf Q}}_n^2.
\end{eqnarray}
Here $\sigma_c$ and $\sigma_{ab}$ are the quasiparticle conductivities
perpendicular and along the layers, respectively.

The Lagrangian and the dissipative function result in the equations of motion for the
phases $\phi_n$ and the vector potential ${\bf A}$
\begin{equation}
\frac{d}{dt}\frac{\delta{\cal L}}{\delta\dot{\phi}_n}-\frac{\delta{\cal
L}}{\delta\phi_n}+ \frac{\delta{\cal R}}{\delta\dot{\phi}_n}=0
\end{equation}
and similar equation for ${\bf A}$. We write them in the form
\begin{eqnarray}
&&\frac{\partial^2}{\partial\tau^2}\left[\varphi_n+\beta\varphi\right]+
(1-\alpha\nabla_n^2)\left(\nu_c\frac{\partial\varphi_n}{\partial\tau}+\sin\varphi_n-\nabla_uh_{y,n}+\nabla_vh_{x,n}\right)
=0, \label{de1}\\
&&(\nabla_n^2-\ell^{-2}T_{ab})h_{y,n}+T_{ab}\nabla_u\varphi_n=0, \\
&&(\nabla_n^2-\ell^{-2}T_{ab})h_{x,n}-T_{ab}\nabla_v\varphi_n=0, \label{de}
\end{eqnarray}
Here we use reduced coordinates $u=x/\lambda_J$ and $v=y/\lambda_J$, reduced
time $\tau=t\omega_p$ and frequency $\omega=\omega_J/\omega_p$ with
$\omega_p=c/(\lambda_c\sqrt{\epsilon_c})$ as well as reduced magnetic field
${\bf h}={\bf B}/B_c$ and
$\beta=NC_s/C_J$. Further, $T_{ab}=1+\nu_{ab}\partial/\partial\tau$, and $C_J=\epsilon_cL_xL_y/(4\pi s)$ is the junction capacitance. We introduce
reduced dissipative parameters
$\nu_{ab}=4\pi\sigma_{ab}/(\gamma^2\epsilon_c\omega_p)$ and
$\nu_c=4\pi\sigma_c/(\epsilon_c\omega_p)$. The shunt capacitor effectively
enhances the capacitance of the Josephson junctions,  but only for synchronized
oscillations; this enhancement is proportional to the number of junctions.

The typical parameters of optimally doped BSCCO at low temperatures
are $\epsilon_c=12$, $s=15.6$ \AA, $\gamma=500$, $\lambda_{ab}=200$
nm, the critical current density
$j_c=\Phi_0c/(8\pi^2s\lambda_c^2)=1700$ A/cm$^2$, $\ell=130$,
$\nu_{ab}=0.2$ and $\nu_c=0.002$, The parameter $\alpha\sim 0.1-1 $.
It was estimated in Ref.~\onlinecite{BK} that with these parameters
crystal optimal for radiation should have sizes $L_z\approx 40$
$\mu$m, $L_x\approx 4-6$ $\mu$m and $L_y\gtrsim 300$ $\mu$m.

The differential equations (\ref{de}) should be completed by the boundary
conditions at $x=\pm L_x/2$ and $y=\pm L_y/2$. Using Eqs.~(\ref{de}) and
continuity of $B_x$ and $B_y$ for the time-independent part of the phase
difference we obtain at these boundaries
\begin{equation}
\nabla_{x,y}\varphi_n=\pm(2\pi s/\Phi_0)B_{y,x}.
\end{equation}
The outside magnetic field is created by the bias DC current and by the induced alternating
current. For $L_y\gg L_x$ we estimate $B_y(x=\pm L_x/2,y=\pm L_y/2)\approx \pm 2\pi
I/(cL_y)$, while $B_x\lesssim B_y$. Here $I=jL_xL_y$ is the total interlayer
bias current and $j$ is the bias current density. Hence, we estimate
time-independent phase difference,
\begin{equation}
\varphi(y=L_y/2)-\varphi(y=-L_y/2)\lesssim
(2\pi)^2sjL_xL_y/(c\Phi_0)\approx 2\pi\sigma_c\omega_JL_xL_y/c^2.
\end{equation}
Here we used the relation $j\approx \sigma_c E_z$ in the resistive state when
voltage is present. The phase difference estimated here is very small for
crystals with dimensions smaller than cm. Neglecting it we use approximation
with $y$-independent phase difference.

For alternating part of the phase difference, we find boundary conditions by
matching electromagnetic fields inside and outside the crystal. \cite{BK}
Outside fields,
in half spaces $|x|>L_x/2$, obey the Maxwell equations, which fix the ratio between electric and magnetic field. Inside the crystal,
${\bf B}_n({\bf r},t)$ given by Eq.~(\ref{de}) and the electric by
Eq.~(\ref{elec}). When $L_y\gg L_x, c/\omega_J$, the predominant radiation is along
the $x$-axis. For weak radiation in the
$y$-direction, we can use the boundary conditions $\nabla_v\varphi_n=0$.
Thus we can omit dependence of $\varphi_n$ on the $y$-coordinate also for
alternating part of the phase differences.
Then we obtain the boundary conditions at $x=\pm L_x/2$:
\begin{equation}
\pm
h_{y,n}(\omega)=\frac{is\ell\omega}{2\sqrt{\epsilon_c}}\sum_m\varphi_m(\omega)[|k_{\omega}|J_0(k_\omega
s|n-m|)+ik_{\omega}N_0(k_{\omega}s|n-m|)], \label{BC}\end{equation} where we
use the Fourier transforms with respect to time, $k_{\omega}=\omega/c$, and
$J_0(x)$ and $N_0(x)$ are the Bessel functions.

The following calculations are similar to those in
Ref.~\onlinecite{BK} but accounting for shunt contribution. We
consider high frequencies $\omega\gg 1$ and $N\gg \ell$ and neglect
finite-size effects along the c-axis. The equation for $y$- and
$n$-uniform phase is
\begin{equation}
\ddot{\varphi}(u)+(\beta/\tilde{L}_x)\int_{-\tilde{L}_x/2}^{\tilde{L}_x/2}
du\ddot{\varphi}(u)+\nu_c\dot{\varphi}+\sin\varphi-\ell^2\nabla_u^2\varphi=0.
\end{equation}
Here $\tilde{L}_x=L_x/\lambda_J$. In the limit $\omega\gg 1$ we look for
solution
\begin{equation}
\varphi(u,\tau)=\omega\tau+\eta(u,\tau), \ \ \ \eta\ll 1.
\end{equation}
The equation for $\eta$ is
\begin{equation}
\ddot{\eta}(u)+(\beta/\tilde{L}_x)\int_{-\tilde{L}_x}^{\tilde{L}_x}
du\ddot{\eta}(u)+\nu_c\dot{\eta}-\ell^2\nabla_u^2\eta=-\sin(\omega\tau),
\end{equation}
which should be solved with the boundary conditions at $u=\pm \tilde{L}_x/2$
\begin{eqnarray}
&&\nabla_u\eta=\pm i\omega\zeta\eta,  \\
&&\zeta=\frac{L_z}{2\ell\sqrt{\epsilon_c}}[|k_{\omega}|-ik_{\omega}{\cal
L}_{\omega}], \ \ \ {\cal L}_{\omega}\approx
\frac{2}{\pi}\ln\left[\frac{5.03}{|k_{\omega}|L_z}\right]. \nonumber
\end{eqnarray}
The solution is
\begin{eqnarray}
&&\eta=(1/2) {\rm Im}[B+A\cos(\overline{k}_{\omega}u)], \ \ \
\overline{k}_{\omega}=\omega/\ell \\
&&B=-[\omega^2(1+\beta\xi)+i\omega\nu_c]^{-1}, \nonumber \\
&&A=i\zeta\{[|\overline{k}_{\omega}\sin(\overline{k}_{\omega}\tilde{L}_x/2)+
i\zeta\omega\cos(\overline{k}_{\omega}\tilde{L}_x/2)][\omega(1+\beta\xi)+i\nu_c]\}^{-1},\nonumber \\
&&\xi=[1+2i\omega\zeta/(\overline{k}_{\omega}^2\tilde{L}_x)]^{-1}. \nonumber
\label{nonun}
\end{eqnarray}
where we approximate
$\sin(\overline{k}_{\omega}\tilde{L}_x/2)\approx \overline{k}_{\omega}\tilde{L}_x/2$.
The first term in $\eta$ is the amplitude of synchronized ($y$- and $n$-independent)
Josephson oscillations. It drops as $\beta$ increases.
The second term describes nonuniform electromagnetic
wave inside the crystal. It is generated at the
boundaries due to radiation field.

To analyze stability of synchronized Josephson oscillations we consider a small
perturbation $\theta_n(u,\tau)$ to the solution uniform in $n$, $\varphi$,
\begin{equation}
\varphi_n(u,\tau)=\varphi(u,\tau)+\theta_n(u,\tau).
\end{equation}
Equations for $\theta_n(u,\tau)$ are obtained by linearization of
Eqs.~(\ref{de}) with respect to $\theta_n(u,\tau)$. The term
$\cos[\eta(\tau)]\theta_n(u,\tau)$ in the linearized equation couples harmonics
with small frequency $\Omega$  to high-frequency terms at $\Omega\pm\omega$. At
$\omega\gg 1$ we can neglect coupling to the higher frequency harmonics
$\Omega\pm m\omega$ with $m>1$ and represent the phase perturbation as
\begin{equation}
\theta_n\approx \sum_q\left[\bar{\theta}_q+\sum_{\pm
}\tilde{\theta}_{q\pm}\exp(\pm
i\omega\tau)\right]\cos(qn)e^{-i\Omega\tau}, \nonumber
\end{equation}
where $q=\pi k/N$, $k=1,2,... N$. The complex eigenfrequencies $\Omega(q)$ are
assumed to be small, $|\Omega|\ll\omega$. We will find them and also conditions
when ${\rm Im}[\Omega]<0$ for all $q$ (stability condition). Substituting
$\theta_n$ into linearized equations (\ref{de1})-(\ref{de}), excluding
oscillating magnetic fields and separating the fast and slow parts, we obtain
for $q\neq 0$  the coupled equations
\begin{eqnarray}
&&\left[\frac{\Omega^2}{1+\alpha_q}+i\nu_c\Omega-\overline{C}\right]\bar{\theta}_q+G_q^{-2}\nabla_u^2\bar{\theta}_q=\frac{\tilde{\theta}_{q+}
+\tilde{\theta}_{q-}}{2}, \\
&&\left[\frac{(\Omega\pm
\omega)^2}{1+\alpha_q}+i\nu_c(\Omega\pm\omega)\right]\tilde{\theta}_{q\pm}+G_{q,\pm}^{-2}\nabla_u^2\tilde{\theta}_{q\pm}=
\frac{\bar{\theta}_q}{2}.\\
&&\overline{C}=\langle\cos\eta\rangle_{\tau}\approx {\rm Re}[\eta_{\omega}]/2\approx -
(1/2){\rm Re}\left[~\frac{1}{(1+\beta\xi)\omega^2+i\omega\nu_c}\right], \\
&&\alpha_q=2\alpha(1-\cos q), \ \ \ G_{q,\pm}^2\approx\frac{2(1-\cos
q)}{1-i(\Omega\mp\omega)\nu_{ab}}+\ell^{-2}.
\end{eqnarray}
Here $G_q=G_{q,0}$. The boundary conditions for slow and fast components at
$u=\pm \tilde{L}_x/2$ and $q\gg\pi/N$ follow from Eq.~(\ref{BC}).

Finally, we obtain Mathieu equation for slow-varying component with $q\neq0$
\begin{eqnarray}
&&\left(\frac{\Omega^2}{1+\alpha_q}+i\nu_c\Omega+\Lambda-V(u)+
G_q^{-2}\nabla_u^2\right)\bar{\theta}_q=0. \\
&&\Lambda={\rm Re}\left[\frac{1}{2[\omega^2(1+\beta\xi)+i\nu_c\omega]}\right]-\frac{1+\alpha_q}{2[\omega^2+\nu_c^2(1+\alpha_q)^2]},
\end{eqnarray}
with potential $V(u)=V_1(u)+V_2(u)$,
\begin{eqnarray}
&&V_1(u)=\frac{1}{2\omega^2}{\rm
Re}\left[\frac{i\zeta\omega\cos(\overline{k}_{\omega}u)}{(1+\beta\xi)[\overline{k}_{\omega}
\sin(\overline{k}_{\omega}\tilde{L}_x/2)
+i\zeta\omega\cos(\overline{k}_{\omega}\tilde{L}_x/2)]}\right], \nonumber\\
&&V_2(u)=\frac{1}{2\omega^2}{\rm Re}\left[\frac{\kappa_+\cos(p_+u)}{(1+\beta\xi)
[p_+\sin(p_+\tilde{L}_x/2)+\kappa_+\cos(p_+\tilde{L}_x/2)]}\right], \nonumber\\
\end{eqnarray}
Here $p_+=\omega G_{q+}$ and $\kappa_+\approx
[(\Omega-\omega)^2G_{q,\beta}^2]/[ (1+\alpha_q)\epsilon_cq\gamma]$.
In the following we consider shunt with moderate capacitance $\beta\leq 1$. In the
lowest order in $\overline{k}_{\omega}\tilde{L}_x=\omega \tilde{L}_x/\ell\ll 1$
the part $V_1(u)$ reduces to a constant, $V_1(u)\approx {\rm Re}[{\cal
K}_{\omega}/(2\omega^2(1+\beta\xi))]$,
\begin{equation}
{\cal K}_{\omega}=\frac{{\cal L}_{\omega}({\cal
L}_{\omega}+a)+1}{({\cal L}_{\omega}+a)^2+1}, \ \ \
a=\frac{\epsilon_c L_x}{L_z}.
\end{equation}
Treating the coordinate-dependent part of $\overline{\theta}_q$ as a small
perturbation, we find expression for $\Omega(q)$,
\begin{eqnarray}
&&\Omega^2+i\nu_c\Omega\approx \frac{1}{2\omega^2}\left\{{\rm Re}\left[\frac{\beta\xi+
{\cal K}_{\omega}}{1+\beta\xi}
+\alpha_q
-\frac{W_2(q)}{1+\beta\xi}\right]\right\},\label{stab} \\
&&W_2(q)=\frac{2}{p_+\tilde{L}_x(p_+/\kappa_++{\rm
cot}(p_+\tilde{L}_x/2))}.
\end{eqnarray}
The first and the second terms in Eq.~(\ref{stab}) represent the stabilization
effects of shunt and radiation, and of the long-range interlayer capacitance, respectively. The last term,
$W_2$, describes the effect of modes $\tilde{\theta}_{q\pm}$ induced inside the crystal due to radiation (parametric excitation of Fiske modes). This term
leads to instability in the limit of zero dissipation and in the absence of
other stabilizing terms. It is much smaller than unity. Capacitive shunt and
radiation introduce the gap in the spectrum of weak distortions and
are most effective in stabilization. Their contributions can both reach order one
for $\epsilon_cL_x<L_z=Ns$
(in the super-radiation regime) and
$NC_s\approx C_J$. As $C_J=60$ cm for $S=1200$ $\mu$m$^2$, it is easy to reach this
condition. In order to achieve the maximal stabilization without sacrificing the radiation power, one needs to choose  $\beta\approx 1$.

We conclude that shunt capacitor stabilizes synchronized oscillations in IJJ
stack.  The effect is particularly useful  in crystals with the small junction
number or at the initial stages of radiation. Shunt may also be used to tune the radiation power.

The work of IM and LNB was carried out under the auspices of the
National Nuclear Security Administration of the U.S. Department of
Energy at Los Alamos National Laboratory under Contract No.
DE-AC52-06NA25396 and supported by the LANL/LDRD Program. The work
of  AK was supported by the U.S. Department of Energy under the
contract No. DE-AC02-06CH11357 (ANL). IM and LNB were supported in
part by the National Science Foundation under Grant No. PHY05-51164.
GBH acknowledges Prof. J. Driscoll of Trinity College, Cambridge,
for supporting this research.

\end{document}